\documentclass[twocolumn]{article}
\usepackage{graphicx} 
\usepackage{titling}
\usepackage{ragged2e} 
\usepackage[numbers,sort&compress,square]{natbib}
\usepackage{hyperref}
\usepackage{geometry} 
\usepackage{amsmath} 
\usepackage{amssymb} 
\usepackage{amsbsy} 
\usepackage{subcaption} 
\usepackage{caption} 
\title{\Large{\textbf{Bayesian Neural Networks  for 2D MRI Segmentation}}}
\author{Lohith Konathala \\ University of Cambridge \\ \vspace{10pt} lk480@cantab.ac.uk}
\date{}

\begin{document}

\maketitle

\begin{center}
    \justifying
    \textbf{\textit{Abstract - }} Uncertainty quantification is vital for safety-critical Deep Learning applications like medical image segmentation. We introduce BA U-Net, an uncertainty-aware model for MRI segmentation that integrates Bayesian Neural Networks with Attention Mechanisms. BA U-Net delivers accurate, interpretable results, crucial for reliable pathology screening. Evaluated on BraTS 2020, our model addresses the critical need for confidence estimation in deep learning-based medical imaging.
\end{center}

\section{Introduction}
Medical image segmentation is a computer vision task that involves extracting regions of interest from a medical image such as an X-ray, MRI or CT scan. Deep Learning models e.g. UNET \citep{unet_ronneberger} have revolutionised the accuracy and speed of automatic segmentation of medical images. This is of great importance as segmentation is often the first stage in an automated diagnostics platform e.g. automatic tumour screening in Brain MRI scans. The quality of the segmentation directly impacts downstream classification tasks hence it is critical that any instances where the model is unable to confidently segment the provided image a medical professional be notified for manual intervention. It is often the case due to variability in medical images that the models will be required to segment scans that are significantly different from their training data (out-of-distribution). Therefore it is necessary for a segmentation model to calculate its confidence in the generated segmentation masks for a given image. This can be done by capturing the aleatoric and epistemic uncertainty \citep{kendall2017uncertainties} in the segmentation masks and calculating thresholds \citep{Seedat_MC_UNet} to prevent erroneous segmentations from being classified. However, frequentist segmentation models such as CNN-based UNETs are incapable of providing these uncertainty estimates and hence it is necessary to turn to Bayesian models for a principled approach to uncertainty estimation. 
\vspace{0.5em}
\newline
A distinguishing factor of a Bayesian approach is modelling the model weights as distributions as opposed to a point estimate. This introduces stochasticity into the model output and allows the training of the model to be regularised by prior knowledge of the weights. Due to the intractability of the posterior distribution \citep{bishop_pattern_rec} it is necessary to approximate the distribution to employ a Bayesian approach. Whilst deterministic approximations do exist, stochastic approximations offer far greater versatility with Markov Chain Monte Carlo (MCMC) and Variational Inference (VI) being the two most well-known approaches. VI-based approaches have been perceived as being computationally expensive and thus alternative approaches have gained more traction such as Deep Ensembles \citep{lakshminarayanan2017simple} and Monte Carlo Dropout \citep{gal2016dropout} which are capable of producing uncertainty estimates but without the computational cost of direct variational inference. In particular, Monte Carlo Dropout Networks have been widely used to estimate uncertainty in segmentation tasks due to their dual purpose as a regularisation mechanism during model training \citep{srivastava2014dropout}. This is despite literature contesting the validity of Dropout as a Bayesian approximation, \citep{osband2016risk}, \citep{folgoc2021mc}.
\\
By leveraging modern probabilistic deep-learning libraries \citep{tran2019bayesian}, we propose a Bayesian UNET that models convolutional kernel weights as distributions and is trained via minimisation of the Evidence Lower Bound (ELBO). Our model is able to out-perform current state-of-the-art approaches for uncertainty-aware MRI segmentation \citep{sagar2021uncertainty} on the BraTS dataset \citep{6975210}, \citep{bakas2019identifying}. 
\subsection*{Contributions}
\begin{itemize}
\item BA U-Net, a novel adaptive segmentation architecture that accurately captures the aleatoric and epistemic uncertainty in semantic segmentation tasks.

\item Achieved high binary segmentation accuracy when segmenting whole tumours (WTs) on the BraTS 2020 dataset with an F1 Score of 0.877 and Mean IoU of 0.788. 

\item Empirically validated model uncertainty estimates by running inference on degraded MRI images - obtained a clear response in the uncertainty metric which correlates with the extent of the degradation.  
\end{itemize}
\section{Related Work}
\subsection{Image Segmentation}
Fully convolutional networks have proven to be highly effective in semantic segmentation tasks \citep{Long_2015_CVPR}. This laid the foundation for the UNET architecture \citep{unet_ronneberger} which proposed an encoder-decoder structure with skip connections between encoding and decoding layers to preserve spatial information that would otherwise be lost during the encoding stage. The UNET architecture has proven to be versatile and has since been adapted to segmentation tasks on a range of imaging modalities. One such adaptation is the inclusion of Attention Gates (Attention-UNET) as proposed by \citep{oktay2018attention}, which was designed specifically for biomedical image segmentation allowing the model to focus on specific target structures. This approach was generalised to a range of vision tasks by \citep{Woo_2018_ECCV} who proposed the Convolutional Block Attention Module (CBAM) that provided both channel and spatial attention mechanisms compared to Attention UNETs which only incorporated a mechanism for spatial attention. The CBAM was successfully applied to UNET by \citep{guo2021sa} to achieve state-of-the-art performance on a Vessel Segmentation task (DRIVE). 
\vspace{0.5em}
\newline
\subsection{Uncertainty Quantification}
Uncertainty can be broadly categorised into two types - aleatoric and epistemic. Aleatoric uncertainty captures the inherent noise in the observations which is invariant to increasing dataset size. Epistemic uncertainty accounts for uncertainty in the model parameters and captures our lack of knowledge regarding the underlying distribution of the data. One of the key contributions of \citep{kendall2017uncertainties} is a unified vision model that can capture both epistemic and aleatoric uncertainty in the model output. This requires a Bayesian Neural Network (BNN) with the model weights drawn from an approximate posterior distribution, $\widehat{W} \sim q(W)$. \\
This results in an output composed of the predictive mean and variance:
\newline
\begin{equation}
    [\mathbf{\hat{y}}, \hat{\sigma}^2] = \mathbf{f}^{\mathbf{\widehat{W}}}(x)
\end{equation}
where $\mathbf{f}$ is a Bayesian CNN with model weights $\mathbf{\widehat{W}}$. 
\citep{kendall2017uncertainties} construct the loss function for this BNN and from this derive an estimator for the predictive variance:
\begin{equation}
    Var(\mathbf{y}) \approx \frac{1}{T} \sum_{t=1}^{T} \mathbf{\hat{y_t}}^2 - \left(\frac{1}{T} \sum_{t=1}^{T} \mathbf{\hat{y_t}} \right)^2 + \frac{1}{T} \sum_{t=1}^{T} \hat{\sigma_t}^2
\end{equation}
\vspace{0.5em}
\newline
\textbf{Monte Carlo Dropout}: Dropout \citep{srivastava2014dropout} is a regularisation technique for neural networks that reduces the dependency of the network on the output of specific neurons. Monte Carlo Dropout \citep{gal2015bayesian} is a clever re-interpretation of dropout offering an efficient way to incorporate uncertainty estimation in a neural network. However, whilst MC Dropout has been widely adopted, the statistical foundation of this approach has been contested. \citep{osband2016risk} argues that MC Dropout gives an approximation for the model risk (known stochasticity) as opposed to the uncertainty (unknown stochasticity). Specifically, \citep{osband2016risk} observed that the variance of the resulting dropout posterior distribution depends only on the dropout rate and the model size with no dependence on the size or observed variance of the dataset.
\vspace{0.5em}
\newline
\textbf{Bayes by Backprop:} An alternative approach to variational inference with BNNs is to use the Bayes by Backprop algorithm \citep{blundell2015weight} which combines variational inference with backpropogation to learn the parameters of the variational posterior. Bayes by Backprop leverages variational learning \citep{graves2011practical} which finds the parameters, $\theta$ of the variational posterior, $ q(\boldsymbol{w} | \theta)$ via minimisation of the KL divergence between the intractable true posterior and the variational approximation. 
\begin{align*}
    \theta^{*} = \min_{\theta} KL[q(\boldsymbol{w}|\theta) || p(\boldsymbol{w} | \mathcal{D})] \\
    \mathcal{F}(\mathcal{D},\theta) = KL[q(\boldsymbol{w}|\theta) || p(\boldsymbol{w})] - \mathbb{E}_{q(\boldsymbol{w}| \theta)}[\log p(\mathcal{D}|\boldsymbol{w})]
\end{align*}
Bayes by Backprop uses a generalisation of the re-parameterisation trick introduced by \citep{kingma2013auto} to re-express the stochastic model weights in terms of a deterministic function, $\boldsymbol{w} = t(\theta, \epsilon)$ where $t(\theta, \epsilon)$ is a deterministic function which enables the derivative of the variational free energy, $\mathcal{F}(\mathcal{D}|\theta)$ to be obtained w.r.t to $\theta = (\mu, \sigma)$.
\vspace{0.5em}
\newline
A key consideration when using re-parameterisation is the variance in the gradient estimates during training which affects training time and convergence. \citep{kingma2015variational} propose the local re-parameterisation trick  improves on the standard re-parameterisation trick used in the Bayes by Backprop algorithm by translating global parameter uncertainty into local uncertainty per data point. As a result of injecting noise, $\epsilon$ locally as opposed to globally the resulting gradient estimator has lower computational cost and reduced variance. However a key limitation of this approach is that it compatible only with fully connected layers with differentiable activation function - making it unsuitable for use with a CNN based architectures such as UNET.
\vspace{0.5em}
\newline
This limitation was addressed by \citep{wen2018} with the Flipout Estimator which offers more versatility being compatible with convolutional, recurrent and fully connected layers. Flipout is also compatible with non-differentiable activation functions such as ReLU which is the activation function of choice for image segmentation models such as UNET \citep{unet_ronneberger}. \newline \citep{wen2018} show that in practice, Flipout achieves a variance reduction comparable with fully independent parameter perturbations $\approx 1/N$, where N is the mini-batch size. Hence for large batch sizes, Flipout achieves a near-zero gradient variance which is the best-case scenario.
\section{Methodology}
\subsection{Model Architecture}
We chose to base our model on the UNET architecture proposed by \citep{unet_ronneberger}. UNETs are fully convolutional networks (FCNs) that are highly versatile and offer strong performance on range of biomedical image segmentation tasks The UNET architecture is capable of being modified with additional features to improve performance on specific segmentation tasks \citep{guo2021sa}, \citep{azad2019bi}.
\newline
\\
To enhance the segmentation performance of the UNET model we include an attention mechanism, Convolutional Block Attention Module \citep{Woo_2018_ECCV}, in the deeper layers within our model. The inclusion of an attention mechanism enables the network to selectively focus on key image features, understand the context of the image features and reduce noise in the final segmentation result. \citep{guo2021sa}, \citep{oktay2018attention} both demonstrate that attention-equipped UNET models achieve superior performance especially in medical segmentation tasks. 
\\
\newline
To incorporate uncertainty estimation capability into the UNET model we chose to replace the deterministic convolutional layers with a variational equivalent - the Variational 2D Convolution Layer equipped the Flipout Gradient Estimator. This implementation of this layer is provided by the TensorFlow Probability Package \citep{dillon2017tensorflow} and uses the Flipout gradient estimator (see Related Work) to estimate the backpropogation gradients.
\\
\newline
The use of variational convolution layers allows for the omission of dropout layers for regularisation. The variational free energy loss function used for training includes a KL divergence term that acts as regularisation by penalising weight distributions that are complex and deviate far from the prior. This helps reduce over-fitting and improves the generalisability of model. 
\subsection{Design Choices}
A fully Bayesian UNET architecture should utilise convolutional layers that represent filter weights as distributions (as opposed to point-estimates) for every layer in the encoder and decoder.
However, in practice the use of Bayesian layers throughout the encoder and decoder significantly comprises the quality of the resulting segmentation and makes the model prohibitively expensive to train due to the added computational cost of using probabilistic layers. In order to maximise the transfer of feature map information from the encoder to the decoder it is advisable to minimise the stochasticity in the encoder. Instead the stochasticity should be introduced in the decoder which is responsible for constructing the segmentation mask for which uncertainty estimates are desired. As a result, we opted for a design which replaced all but the last two convolutional layers of the decoder with the Convolutional 2D Layer with Flipout Gradient Estimator. The decision to retain deterministic convolution layers for the final stage of the network was motivated by the potential for imprecise segmentations due to Bayesian layers adding uncertainty in the final layer weights. This is desirable in deeper model layers as this will ensure sensitive uncertainty estimates for the final segmentation however this cannot be at the cost of segmentation accuracy. 
\begin{figure}[hbt!]
    \centering
    \includegraphics[scale=0.15]{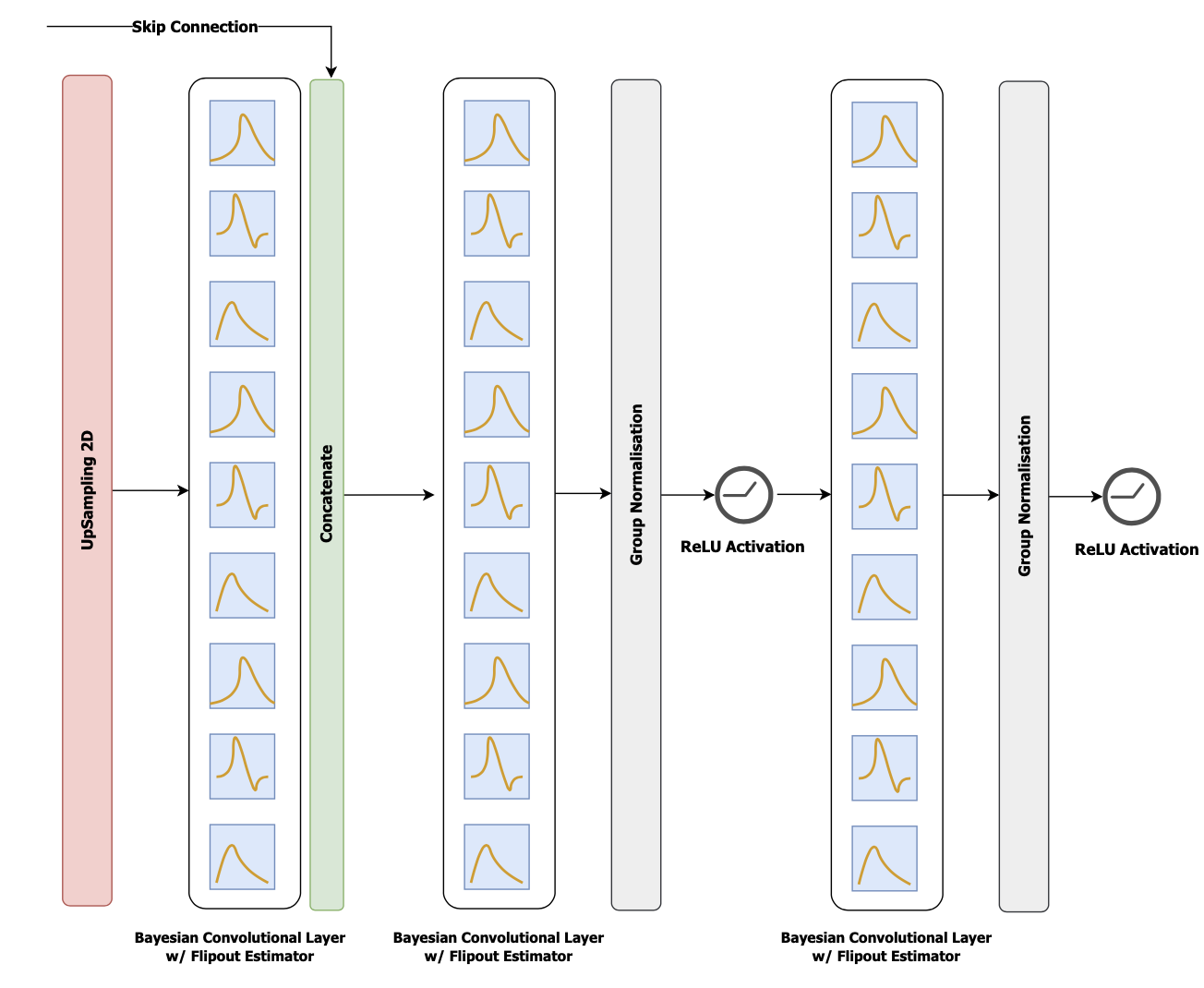}
    \caption{Bayesian Attention U-Net Decoder Block}
    \label{fig:enter-label}
\end{figure}
\newline
Having established the optimal configuration of the decoder, we the investigated the impact of varying the number of filters used to initialise the model. For each stage within the encoder the number of filters doubles until the bottleneck stage is reached where the number of filters is $2^n$ (n - number of encoder stages) times the number of filters used in the first stage. Each stage is made up of two matched convolutional layers separated by a non-linear activation and normalisation layer. The UNET architecture proposed by \citep{unet_ronneberger} utilised 64 filters in the first stage leading to 1024 filters in the bottleneck. However, this would result in a model with over 50 million parameters due to the use of probabilistic layers which is too computationally expensive to train given our limited compute resources. As a result it was necessary to reduce the number of filters at each convolutional layer. Table 1 shows the performance of our model when initialising with 16 $\&$ 32 filters.
\begin{table}[h]
\centering
\begin{tabular}{|c|c|c|c|}
\hline
\textbf{Filters} & \textbf{F1} & \textbf{IoU} & \textbf{Parameters}\\ \hline
16 & 0.836 & 0.741 & 3,181,809 \\ \hline
32 & 0.865 & 0.779  & 12,717,537 \\ \hline
\end{tabular}
\caption{Effect of varying filter initialisation on model accuracy and complexity}
\label{your-table-label}
\end{table}
\noindent
Table 1 clearly shows that despite the added computational cost of using 32 filters at initialisation, the segmentation accuracy of the model increases (3.5$\%$ increase in F1 Score and 5.1$\%$ increase in IoU). As a result, we opted to initialise our model with 32 filters and used a batch size of 32 to ensure we remain within the constraints of our available GPU RAM.  
\\
\newline
In addition to the use of Bayesian layers in the decoder, our model leverages attention mechanisms to improve the segmentation accuracy.  To empirically determine the optimal configuration of attention mechanisms in our model we compared the training fit and segmentation accuracy (F1 Score $\&$ IoU) of 2 different configurations. The first configuration was consistent with \citep{guo2021sa} where we place the Convolutional Block Attention Module (CBAM) proposed by \citep{wen2018} in the bottleneck stage. The bottleneck stage contains the most abstract and compressed form of the input image and by applying the channel and spatial attention mechanisms the model is able to selectively focus relevant features and discern important contextual information. The second configuration aimed to build-on the advantages of including the CBAM at the bottleneck by also including the module in the final layer of the encoder (Encoder Layer 4) and first layer of the decoder (Decoder Layer 1). The rationale  behind an attention mechanism in the final encoder stage was refinement of feature representations and selective emphasis of relevant features prior to entering the bottlenecks. On the other hand, including an attention mechanism in the first decoder layer may guide the network when constructing the segmentation mask and ensure that the features provided via skip connections from the encoder are attentively refined.  
\begin{table}[h]
\centering
\begin{tabular}{|c|c|c|c|}
\hline
\textbf{Location} & \textbf{F1 } & \textbf{IoU} \\ \hline
Bottleneck Only & 0.877 & 0.792  \\ \hline
Central Layers & 0.876 & 0.788\\ \hline
\end{tabular}
\caption{Effect of varying filter initialisation on model accuracy and complexity}
\label{your-table-label}
\end{table}
\noindent
We can see from Table 2, that we achieved similar performance when incorporating the attention module into the central layers of our model as including it solely in the bottleneck stage. In fact, inclusion of the attention module in the bottleneck alone resulted in better accuracy on our test data. This was contrary to our initial hypothesis that the inclusion of an attention modules in additional layers would increase segmentation accuracy. A possible explanation for the slight reduction in performance is the the disruption of the feature extraction process in the bottleneck by allowing selective focusing in adjacent layers. As a result of our testing, we opted to use a single attention module located in the bottleneck which has the added benefit of reduced model complexity. 
\\
\newline
Finally, in order to improve training efficiency we included feature normalisation which maintains gradient flow during backpropogation (mitigates against vanishing $\&$ exploding gradients) and enables higher learning rates to speed up the training process. In particular, Batch Normalisation and Group Normalisation are often used for UNet-based architectures. Batch normalisation normalises the output of the activation layer by subtracting the batch mean and dividing by the batch standard deviation. However this results in a dependency on the batch size with small batch sizes resulting in noisy mean and standard deviation estimates that are not representative of the entire dataset. Group normalisation divides channels into groups and computes group-specific mean and variance and therefore is independent of the batch size. However, this results in the addition of a hyper-parameter to control the number of groups which will require tuning for a particular segmentation task. We compared the performance of both normalisation techniques and observed that Group Normalisation with a group size of 32 achieved an 8.8 $\%$ higher F1 Score and 8.2$\%$ higher IoU and as a result we opted to used Group Normalisation for our final model architecture. 
\\
\newline
Figure 3 shows the architecture of the BA U-Net incorporating all of the design choices which we have presented in this section and justified using empirical results. 
\begin{figure}[hbt!]
    \centering
    \includegraphics[scale=0.25]{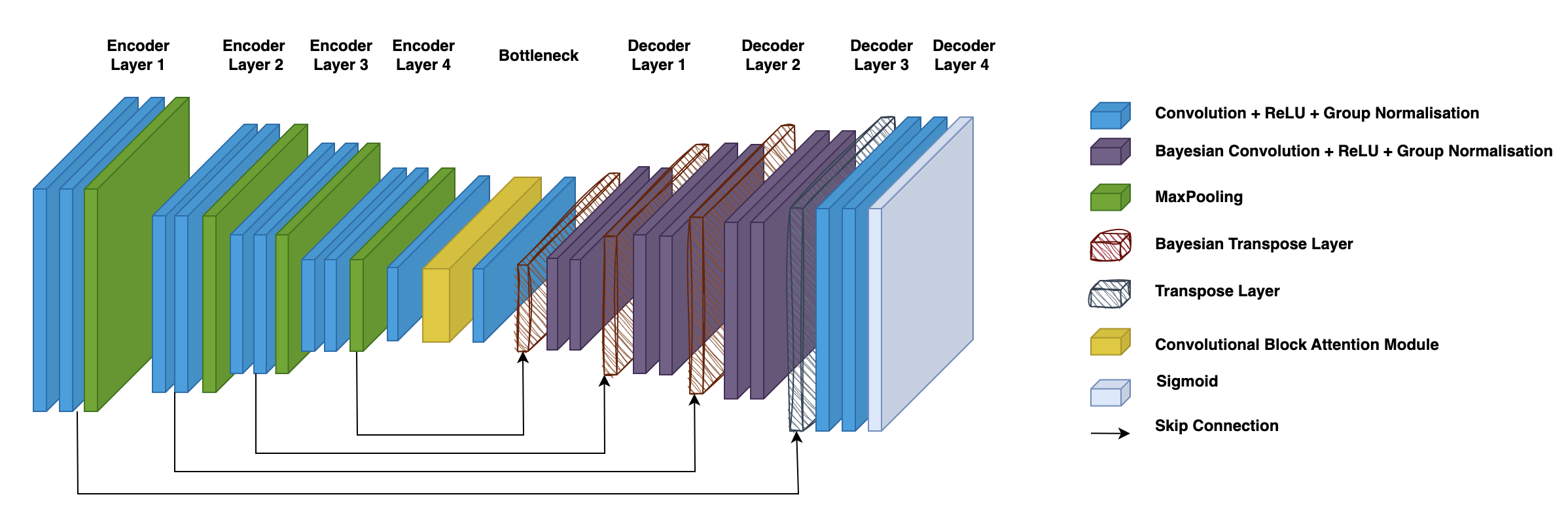}
    \caption{Architecture of Bayesian Attention U-Net}
    \label{fig:enter-label}
\end{figure}
\section{Experimental Setup}
\subsection{Dataset}
To validate the performance our of proposed architecture we chose to use the publicly available BraTS 2020 dataset \citep{bakas2019identifying}, \citep{6975210}. The dataset consists of multi-institutional pre-operative multi-modal MRI scans that were clinically-acquired. It includes scans of glioblastoma (HGG) and lower grade glioma (LGG). The MRI data is multi-modal and contains four-types of MRI sequences: Native (T1), Post-Contrast T1-weighted, T2-weighted and T2 Fluid Attenuated Inversion Recovery (FLAIR). The BraTS 2020 dataset contains annotated multi-modal MRI data for 369 diffuse glioma patients with the ground truth tumour labels provided by expert human annotators. Each of the MRI volume is skull-stripped and co-aligned to the SRI24 anatomical atlas \citep{rohlfing2010sri24}.
\begin{figure}[hbt!]
    \centering
    \includegraphics[scale=0.4]{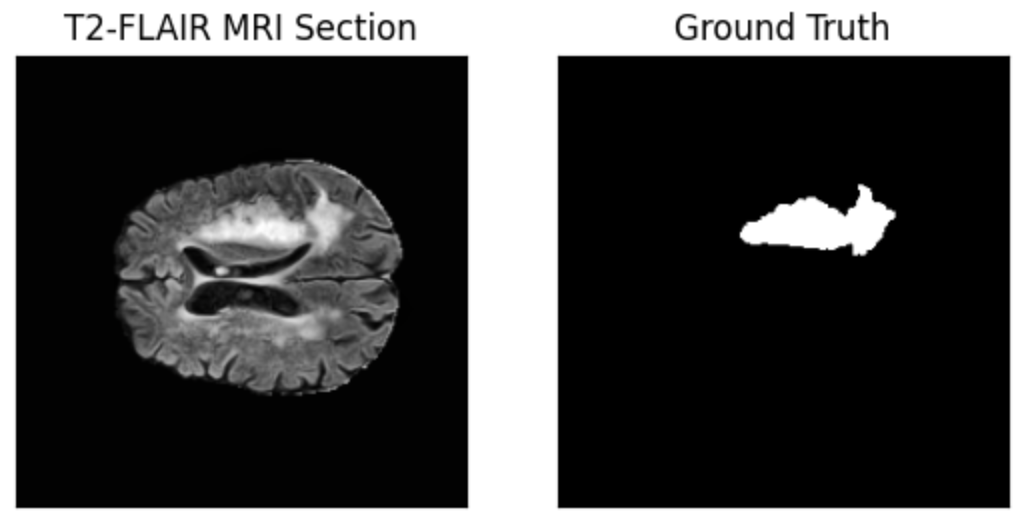}
    \caption{BraTS 2020 T2-FLAIR MRI section with ground-truth tumour annotation}
    \label{fig:enter-label}
\end{figure}
In order to train our segmentation model it is necessary to process this data further and create a training, validation and test set. Our segmentation model requires 2D sections with an associated ground-truth segmentation (as shown in Figure 4). Each MRI volume contains 155 slices however slices from at the extremities of the volume are not of interest so instead we sample 100 slices from each volume to improve the data processing efficiency. As mentioned above each MRI volume contains four sequence modalities of which T2-FLAIR is optimal for visualising fluid-filled structures in the brain such as peritumoral edema. This results a clearer view of the tumor's boundaries and surrounding affected tissue in a 2D section. 
\\
\newline 
To construct our training, validation and test datasets we randomly selected 2000 MRI sections from the processed data and split this into our three datasets according to a 70:10:20  ratio. This yielded 1700 training images, 100 validation images and 200 test images each with corresponding ground-truth annotations. 
\subsection{Model Training}
Our segmentation model was trained using the Bayes by Backprop algorithm proposed by \citep{blundell2015weight} which requires the use of the variational free energy, $\mathcal{F}(\mathcal{D},\theta)$ as the loss function. However using the amendment proposed  \citep{graves2011practical} we can adapt this function to be compatible with a mini-batch gradient descent scheme. Consider randomly splitting the dataset $\mathcal{D}$ into M subsets i.e. $\mathcal{D}_1, \mathcal{D}_2...\mathcal{D}_M$, the mini-batch cost, $\mathcal{F}(\mathcal{D}_i,\theta)$ for $i \in {1,2,...,M}$ for each epoch, E is given by:
\begin{equation}
    \mathcal{F}^E(\mathcal{D}_i,\theta) = \frac{1}{M} KL[q(\boldsymbol{w}|\theta) || p(\boldsymbol{w})] - \mathbb{E}_{q(\boldsymbol{w}| \theta)}[\log p(\mathcal{D}_i|\boldsymbol{w})]
\end{equation}
This is clearly equivalent to the variational free energy function defined in Equation 6 as $\sum_{i=1}^{M} \mathcal{F}(\mathcal{D}_i,\theta) = \mathcal{F}(\mathcal{D}, \theta)$. 
\\
\newline
The derivation of the variational free energy function, $\mathcal{F}(\mathcal{D}_i,\theta)$ shows using this as a cost function represents a trade-off between the likelihood cost and the KL divergence penalty.
\\
\newline
Incorporating these amendments gives us the final form for our variational free energy loss function:
\begin{equation}
        \mathcal{F}^E(\mathcal{D}_i,\theta) = \beta_0 \pi_i KL[q(\boldsymbol{w}|\theta) || p(\boldsymbol{w})] - \mathbb{E}_{q(\boldsymbol{w}| \theta)}[\log p(\mathcal{D}_i|\boldsymbol{w})]
\end{equation}
We opted for the Adam stochastic optimiser \citep{kingma2014adam} with an intial learning rate, $\alpha=0.001$. A ReduceLROnPlateau callback was used to reduce the learning rate if the loss function plateaus for more than 10 epochs. Our model was trained on an NVIDIA A100 GPU with 40GB of VRAM provided through Google Colab. Our training images (2D MRI T2-FLAIR Sections) were re-sized to 256x256x1 with a maximum batch size of 32. The model was trained over 15 epochs and achieved a 99.2$\%$ training accuracy. 
\subsection{Model Inference}
To obtain the mean and variance of the predictive distribution, $\mathbf{f}^{\mathbf{\widehat{W}}}(x)$, we can use Monte Carlo sampling. This is equivalent to running $T$ forward passes for the same input image and estimating the mean, $\mathbf{\hat{y}}$  and variance, $\hat{\sigma}^2$  from the set of output segmentation. From these estimates we can compute the aleatoric and epistemic uncertainty by utilising the predictive variance estimator derived by \citep{kendall2017uncertainties}:
\begin{equation}
    \underbrace{ \frac{1}{T} \sum_{t=1}^{T} \mathbf{\hat{y_t}}^2 - \left(\frac{1}{T} \sum_{t=1}^{T} \mathbf{\hat{y_t}} \right)^2 }_{\text{Aleatoric Uncertainty}} + \underbrace{\frac{1}{T} \sum_{t=1}^{T} \hat{\sigma_t}^2}_{\text{Epistemic Uncertainty}}
\end{equation}
The value of $T$ used can be determined empirically by observing how the variance of the aleatoric and epistemic uncertainty estimates for validation images changes with an increasing number of MC samples. Figure 5 (below) shows the results from varying the number of MC samples from 5 to 30. It should be noted that increasing the number of MC samples increases the computational cost at inference time as a result we opted for $T=20$ which results in adequate variance reduction whilst maintaining efficiency during inference. 
\begin{figure}[hbt!]
   \begin{minipage}{0.5\textwidth}
     \centering
     \includegraphics[width=.7\linewidth]{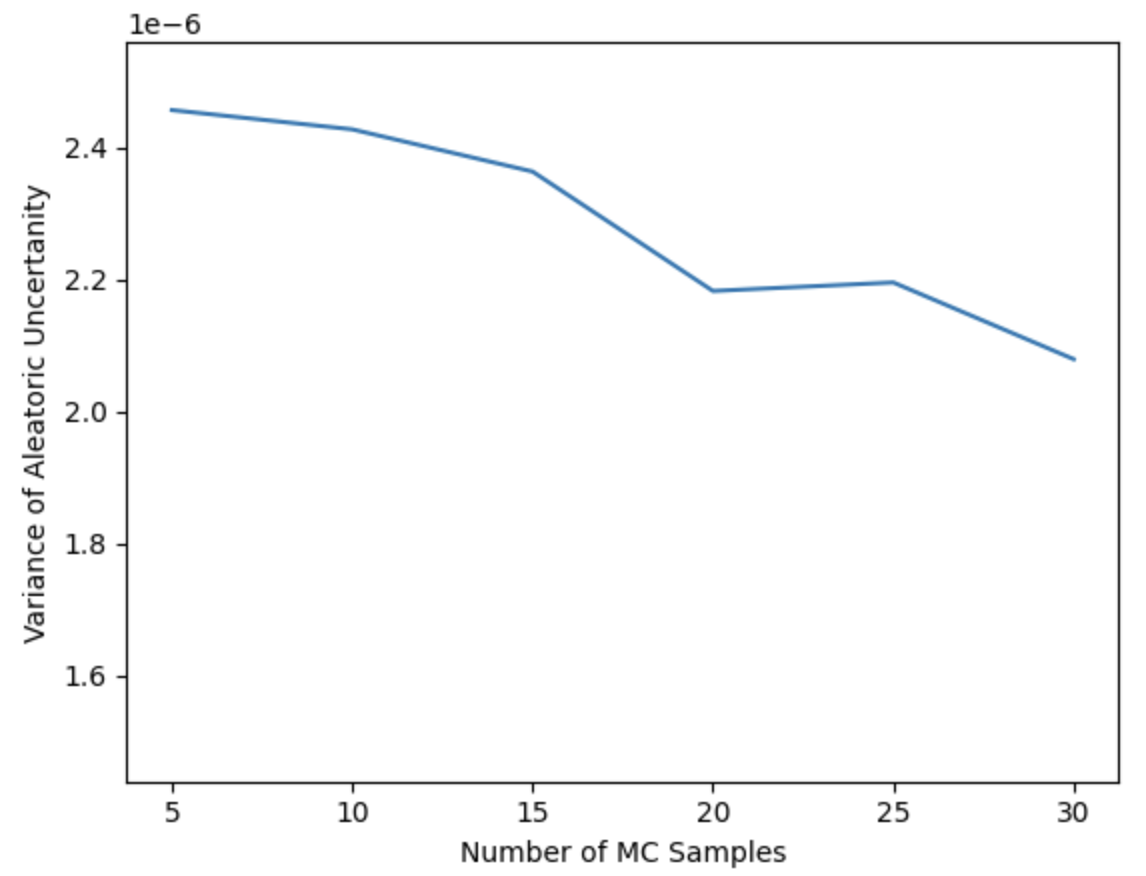}
   \end{minipage}\hfill
   \begin{minipage}{0.5\textwidth}
     \centering
     \includegraphics[width=.7\linewidth]{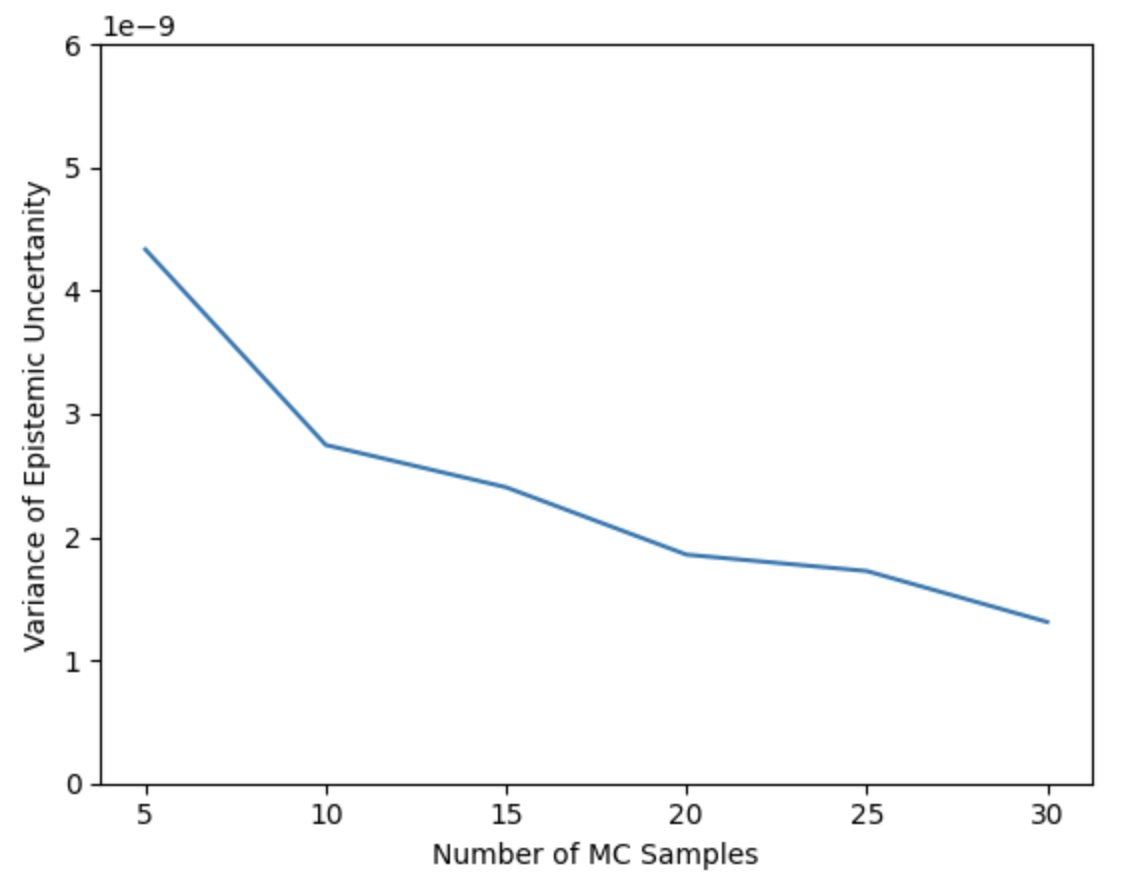}
   \end{minipage}
   \caption{Variance Analysis of Aleatoric and Epistemic Uncertainties}
\end{figure}
\newpage
\noindent
The mean $\mathbf{\hat{y}}$ is used as the MRI segmentation and its accuracy is measured using two accuracy metrics, F1 Score and Intersection over Union (IoU).  F1 score provides a single measurement capturing both model precision and recall (sensitivity). By considering FP and FN results F1 scores are robust to imbalances in the dataset unlike simpler metrics such as binary accuracy. However, an F1 score is not robust to variations in target object size which is a key feature of medical images as a result we opted to use an additional accuracy metric -  Intersection over Union (IoU). The IoU provides a quantitative measure of a model's ability to localise and object within an image and is robust to variations in object size. IoU is also well-suited to scenes where objects may be occluded.
\\
\newline
Having obtained estimates of the $\mathbf{\hat{y}}$  and variance, $\hat{\sigma}^2$ via Monte Carlo sampling we can use the formulation proposed by \citep{kendall2017uncertainties} to generate 2D maps of the aleatoric and epistemic uncertainty. We can calculate the mean of these 2D uncertainty maps to get an estimate for the average aleatoric and epistemic uncertainty in the segmentation of a given image. Combining the average aleatoric and epistemic uncertainties gives us an estimate for the overall predictive variance in the final segmentation as shown in Equation 2.
\section{Results}
In this section we present the inference results of our BA U-Net segmentation model on the BraTS 2020 dataset. We include the the predicted segmentation and a visualisation of the aleatoric and epistemic uncertainty.  
\begin{figure}
    \centering
    \begin{minipage}{\textwidth}
        \includegraphics[width=0.4\textwidth]{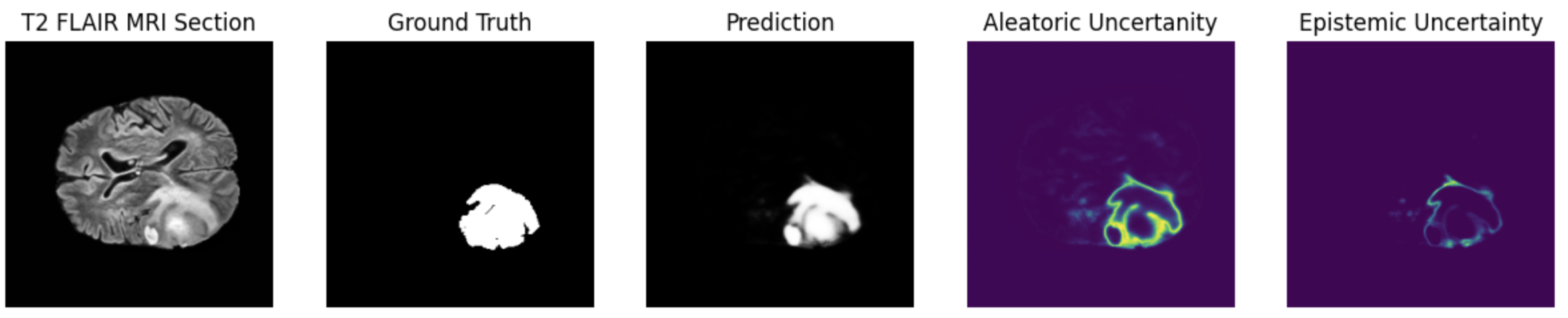}
    \end{minipage}\\ 
    \begin{minipage}{\textwidth}
        \includegraphics[width=0.4\textwidth]{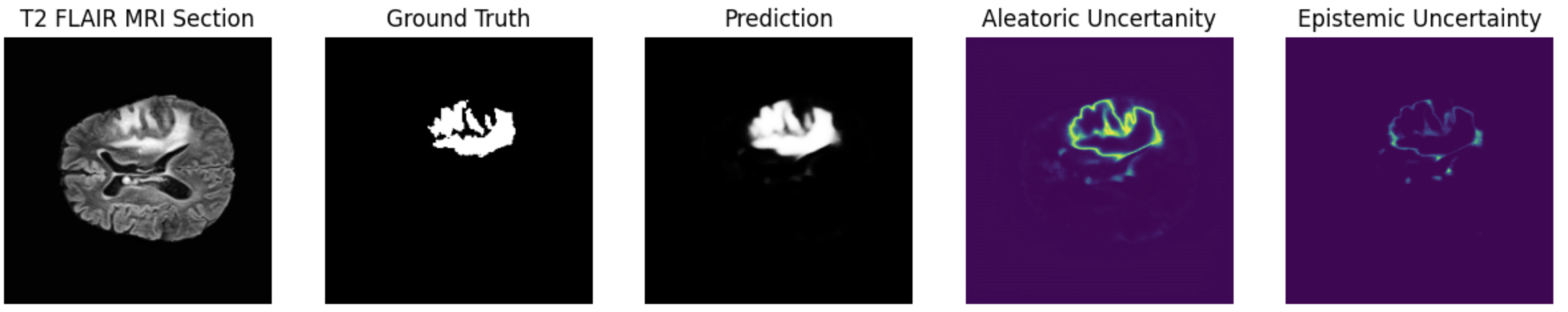}
    \end{minipage}\\
    \begin{minipage}{\textwidth}
        \includegraphics[width=0.4\textwidth]{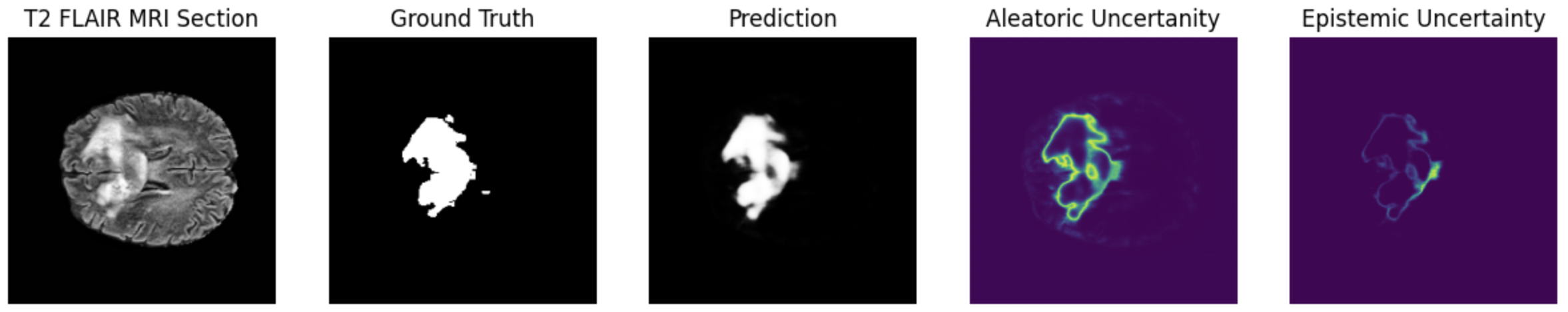}
    \end{minipage}
    \caption{BA U-Net Inference on BraTS 2020 Dataset}
\end{figure}
Our test dataset contained 314 images and we achieved an average \textbf{F1 Score: 0.877} and an \textbf{IoU: 0.792}. This represents an improvement over existing variational approaches to MRI image segmentation \citep{sagar2021uncertainty} and is competitive when compared with submissions to the BraTS 2020 Competition.  
For each segmentation we were able to estimate the aleatoric and epistemic uncertainty however validation of these uncertainties is challenging due to the lack of standardised benchmarks. 
\\
\newline
As a result, we opted to test the performs of our approach on progressively degraded MRIs paying close attention to the response of the overall model uncertainty (aleatoric + epistemic). We measured the response to three types of image degradation: Gaussian Blur, Rician Noise and Brightness/Contrast Changes. These were chosen on the basis they best reflect the alterations that are commonly found in real-world medical image data. 
\subsection{Gaussian Blur}
\begin{equation}
    G_{\sigma}(x,y) = \frac{1}{2\pi \sigma^2} exp(-\frac{x^2 + y^2}{2 \sigma^2}) 
\end{equation}
Equation 14 is a 2D Gaussian kernel and we apply this function to our MRI image prior to inference using our uncertainty-aware model. Figure 7 below shows our results for $\sigma = 2.6$ and $\sigma = 4.4$. 
\begin{figure}[!htb]
    \centering
    \includegraphics[width=0.98\columnwidth]{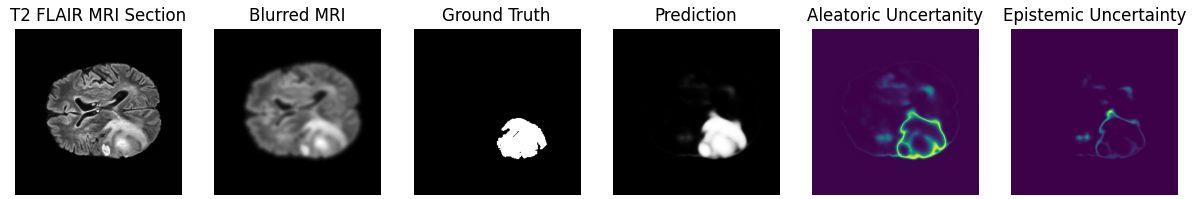} 
    \label{fig:firstimage}
    \includegraphics[width=0.98\columnwidth]{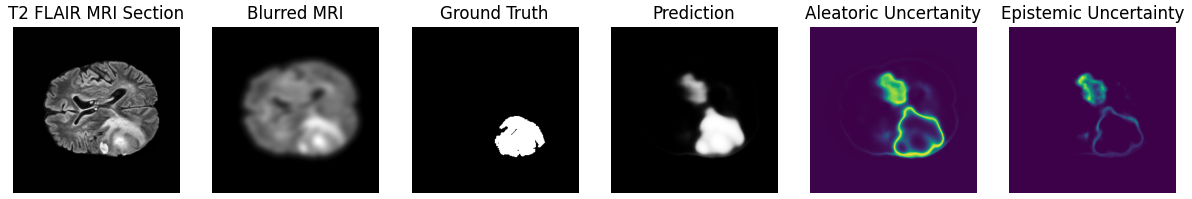} 
    \label{fig:secondimage}
    \caption{Inference on Gaussian Blur MRI}
\end{figure}

\subsection{Rician Noise}
The noise present in magnitude MRI images has been shown to be governed by a Rician distribution \citep{pmid8598820}. A Rician distribution can be approximated using the Rayleigh distribution in regions where there is no NMR signal. As a result, we apply noise governed by this Rayleigh distribution. 
\begin{equation}
f(x; \sigma) = \frac{x}{\sigma^2} e^{-x^2 / (2\sigma^2)}, \quad x \geq 0
\end{equation}
\noindent
where \( \sigma \) is the scale parameter of the distribution.
\noindent
The Rayleigh distribution is the magnitude of a two-dimensional vector whose components are independent Gaussian random variables with equal variances and zero means. If \( X \) and \( Y \) are such Gaussian random variables, then the random variable representing their magnitude, \( Z \) will be Rayleigh distributed. 
\begin{equation}
Z = \sqrt{X^2 + Y^2}\hspace{1em}  X \sim \mathcal{N}(0, \sigma^2), \quad Y \sim \mathcal{N}(0, \sigma^2)
\end{equation}
Figure 8 below shows our results for $\sigma = 0.3$
\begin{figure}[hbt!]
    \centering
    \includegraphics[scale=0.25]{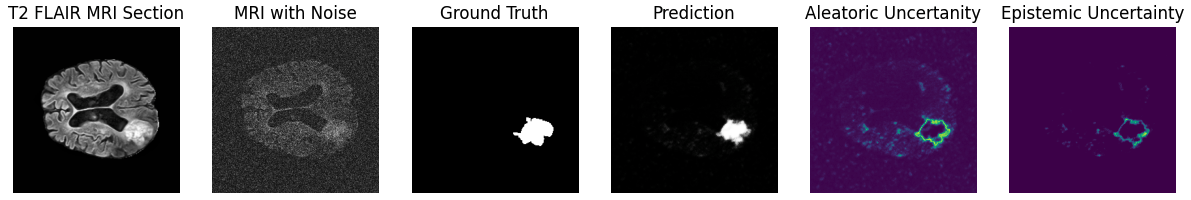}
    \caption{Inference on Rician Noise Degraded MRI}
    \label{fig:enter-label}
\end{figure}
\section{Discussion}
The results presented in the previous section were obtained by conducting 20 stochastic forward passes on a randomly selected test image. The prediction mean represents the segmentation, while model uncertainty was calculated using Equation (2). Model uncertainty comprises both aleatoric and epistemic components. Aleatoric uncertainty reflects inherent data variability, while epistemic uncertainty stems from model limitations.
\\
\newline
Figure 6 illustrates results on clean MRI sections, visualizing both uncertainty components. The model demonstrates high confidence in its predicted segmentation, with elevated uncertainty concentrated at tumor boundaries. This provides interpretability unavailable in deterministic models, even for clear MRI scans. In Figure 6a, despite a high F1 Score (0.924), the model exhibits uncertainty regarding a potential sub-structure within the tumor, highlighted in the aleatoric uncertainty visualization.
\\
\newline
To further evaluate our BA U-Net's capabilities, we tested its inference on degraded MRI scans. We applied three types of image degradation: Gaussian Blur, Rician Noise, and Brightness/Contrast Changes. These degradations are representative of real-world MRI data imperfections, and our approach's applicability hinges on demonstrating increased model uncertainty in such cases.
\newline
\\
Figure 7 presents inference results on Gaussian Blur-degraded MRI sections for two standard deviation ($\sigma$) values. At $\sigma = 2.6$ (Figure 7a), the slight effect on image quality is reflected in a marginal increase in model uncertainty. The uncertainty visualization reveals model indecision regarding a potential second structure within the MRI. This provides precise insight into weak Gaussian blur effects on MRI images and resultant segmentation errors. Figure 7b shows results for $\sigma = 4.4$, where significant blurring led to a 61\% increase in model uncertainty. The predicted segmentation erroneously includes a second tumor, but the uncertainty visualization highlights the model's low confidence in this prediction.
\newline
\\
Following Gaussian Blur, we applied Rician noise approximation, chosen on the basis that it effectively models noise distribution in magnitude MRI images. Figure 8 demonstrates a 21.5\% increase in model uncertainty for noisy MRI images. Uncertainty visualizations show that noise interfered with model confidence in regions typically considered highly certain in clean MRI images.
\newpage
\section{Conclusion}
In conclusion, we present BA U-Net, an uncertainty-aware segmentation model that achieves competitive performance on the BraTS 2020 whole tumor segmentation task with an F1 score of 0.877 and IoU of 0.792. Our model's key advantage lies in its ability to provide credible uncertainty estimates, validated through systematic MRI degradation experiments. BA U-Net demonstrates increased uncertainty in response to image artifacts, offering valuable insights into potential mis-classifications. This approach not only delivers accurate tumor segmentations but also provides critical uncertainty information, enhancing reliability in clinical settings where image quality may vary. Our work represents a significant step towards more robust and trustworthy AI-assisted brain tumor segmentation.
\newpage
\bibliographystyle{IEEEtran}
\bibliography{biblio.bib}

\begin{thebibliography}{10}
\providecommand{\url}[1]{#1}
\csname url@samestyle\endcsname
\providecommand{\newblock}{\relax}
\providecommand{\bibinfo}[2]{#2}
\providecommand{\BIBentrySTDinterwordspacing}{\spaceskip=0pt\relax}
\providecommand{\BIBentryALTinterwordstretchfactor}{4}
\providecommand{\BIBentryALTinterwordspacing}{\spaceskip=\fontdimen2\font plus
\BIBentryALTinterwordstretchfactor\fontdimen3\font minus \fontdimen4\font\relax}
\providecommand{\BIBforeignlanguage}[2]{{%
\expandafter\ifx\csname l@#1\endcsname\relax
\typeout{** WARNING: IEEEtran.bst: No hyphenation pattern has been}%
\typeout{** loaded for the language `#1'. Using the pattern for}%
\typeout{** the default language instead.}%
\else
\language=\csname l@#1\endcsname
\fi
#2}}
\providecommand{\BIBdecl}{\relax}
\BIBdecl

\bibitem{unet_ronneberger}
O.~Ronneberger, P.~Fischer, and T.~Brox, ``U-net: Convolutional networks for biomedical image segmentation,'' in \emph{Medical Image Computing and Computer-Assisted Intervention -- MICCAI 2015}, N.~Navab, J.~Hornegger, W.~M. Wells, and A.~F. Frangi, Eds.\hskip 1em plus 0.5em minus 0.4em\relax Cham: Springer International Publishing, 2015, pp. 234--241.

\bibitem{kendall2017uncertainties}
A.~Kendall and Y.~Gal, ``What uncertainties do we need in bayesian deep learning for computer vision?'' \emph{Advances in neural information processing systems}, vol.~30, 2017.

\bibitem{Seedat_MC_UNet}
\BIBentryALTinterwordspacing
N.~Seedat, ``Mcu-net: {A} framework towards uncertainty representations for decision support system patient referrals in healthcare contexts,'' \emph{CoRR}, vol. abs/2007.03995, 2020. [Online]. Available: \url{https://arxiv.org/abs/2007.03995}
\BIBentrySTDinterwordspacing

\bibitem{bishop_pattern_rec}
C.~M. Bishop, \emph{Pattern Recognition and Machine Learning (Information Science and Statistics)}.\hskip 1em plus 0.5em minus 0.4em\relax Berlin, Heidelberg: Springer-Verlag, 2006.

\bibitem{lakshminarayanan2017simple}
B.~Lakshminarayanan, A.~Pritzel, and C.~Blundell, ``Simple and scalable predictive uncertainty estimation using deep ensembles,'' \emph{Advances in neural information processing systems}, vol.~30, 2017.

\bibitem{gal2016dropout}
Y.~Gal and Z.~Ghahramani, ``Dropout as a bayesian approximation: Representing model uncertainty in deep learning,'' in \emph{international conference on machine learning}.\hskip 1em plus 0.5em minus 0.4em\relax PMLR, 2016, pp. 1050--1059.

\bibitem{srivastava2014dropout}
N.~Srivastava, G.~Hinton, A.~Krizhevsky, I.~Sutskever, and R.~Salakhutdinov, ``Dropout: a simple way to prevent neural networks from overfitting,'' \emph{The journal of machine learning research}, vol.~15, no.~1, pp. 1929--1958, 2014.

\bibitem{osband2016risk}
I.~Osband, ``Risk versus uncertainty in deep learning: Bayes, bootstrap and the dangers of dropout,'' in \emph{NIPS workshop on bayesian deep learning}, vol. 192, 2016.

\bibitem{folgoc2021mc}
L.~L. Folgoc, V.~Baltatzis, S.~Desai, A.~Devaraj, S.~Ellis, O.~E.~M. Manzanera, A.~Nair, H.~Qiu, J.~Schnabel, and B.~Glocker, ``Is mc dropout bayesian?'' \emph{arXiv preprint arXiv:2110.04286}, 2021.

\bibitem{tran2019bayesian}
D.~Tran, M.~Dusenberry, M.~Van Der~Wilk, and D.~Hafner, ``Bayesian layers: A module for neural network uncertainty,'' \emph{Advances in neural information processing systems}, vol.~32, 2019.

\bibitem{sagar2021uncertainty}
A.~Sagar, ``Uncertainty quantification using variational inference for biomedical image segmentation,'' 2021.

\bibitem{6975210}
B.~H. Menze, A.~Jakab, S.~Bauer, J.~Kalpathy-Cramer, and F.~et~al., ``The multimodal brain tumor image segmentation benchmark (brats),'' \emph{IEEE Transactions on Medical Imaging}, vol.~34, no.~10, pp. 1993--2024, 2015.

\bibitem{bakas2019identifying}
S.~Bakas, M.~Reyes, A.~Jakab, S.~Bauer, M.~Rempfler, and A.~C. et~al., ``Identifying the best machine learning algorithms for brain tumor segmentation, progression assessment, and overall survival prediction in the brats challenge,'' 2019.

\bibitem{Long_2015_CVPR}
J.~Long, E.~Shelhamer, and T.~Darrell, ``Fully convolutional networks for semantic segmentation,'' in \emph{Proceedings of the IEEE Conference on Computer Vision and Pattern Recognition (CVPR)}, 2015.

\bibitem{oktay2018attention}
O.~Oktay, J.~Schlemper, L.~L. Folgoc, M.~Lee, M.~Heinrich, K.~Misawa, K.~Mori, S.~McDonagh, N.~Y. Hammerla, B.~Kainz \emph{et~al.}, ``Attention u-net: Learning where to look for the pancreas,'' \emph{arXiv preprint arXiv:1804.03999}, 2018.

\bibitem{Woo_2018_ECCV}
S.~Woo, J.~Park, J.-Y. Lee, and I.~S. Kweon, ``Cbam: Convolutional block attention module,'' in \emph{Proceedings of the European Conference on Computer Vision (ECCV)}, 2018.

\bibitem{guo2021sa}
C.~Guo, M.~Szemenyei, Y.~Yi, W.~Wang, B.~Chen, and C.~Fan, ``Sa-unet: Spatial attention u-net for retinal vessel segmentation,'' in \emph{2020 25th international conference on pattern recognition (ICPR)}.\hskip 1em plus 0.5em minus 0.4em\relax IEEE, 2021, pp. 1236--1242.

\bibitem{gal2015bayesian}
Y.~Gal and Z.~Ghahramani, ``Bayesian convolutional neural networks with bernoulli approximate variational inference,'' \emph{arXiv preprint arXiv:1506.02158}, 2015.

\bibitem{blundell2015weight}
C.~Blundell, J.~Cornebise, K.~Kavukcuoglu, and D.~Wierstra, ``Weight uncertainty in neural network,'' in \emph{International conference on machine learning}.\hskip 1em plus 0.5em minus 0.4em\relax PMLR, 2015, pp. 1613--1622.

\bibitem{graves2011practical}
A.~Graves, ``Practical variational inference for neural networks,'' \emph{Advances in neural information processing systems}, vol.~24, 2011.

\bibitem{kingma2013auto}
Kingma and Welling, ``Auto-encoding variational bayes,'' \emph{arXiv preprint arXiv:1312.6114}, 2013.

\bibitem{kingma2015variational}
D.~P. Kingma, T.~Salimans, and M.~Welling, ``Variational dropout and the local reparameterization trick,'' \emph{Advances in neural information processing systems}, vol.~28, 2015.

\bibitem{wen2018}
\BIBentryALTinterwordspacing
Y.~Wen, P.~Vicol, J.~Ba, D.~Tran, and R.~B. Grosse, ``Flipout: Efficient pseudo-independent weight perturbations on mini-batches,'' \emph{CoRR}, vol. abs/1803.04386, 2018. [Online]. Available: \url{http://arxiv.org/abs/1803.04386}
\BIBentrySTDinterwordspacing

\bibitem{azad2019bi}
R.~Azad, M.~Asadi-Aghbolaghi, M.~Fathy, and S.~Escalera, ``Bi-directional convlstm u-net with densley connected convolutions,'' in \emph{Proceedings of the IEEE/CVF international conference on computer vision workshops}, 2019, pp. 0--0.

\bibitem{dillon2017tensorflow}
J.~V. Dillon, I.~Langmore, D.~Tran, E.~Brevdo, S.~Vasudevan, D.~Moore, B.~Patton, A.~Alemi, M.~Hoffman, and R.~A. Saurous, ``Tensorflow distributions,'' \emph{arXiv preprint arXiv:1711.10604}, 2017.

\bibitem{rohlfing2010sri24}
T.~Rohlfing, N.~M. Zahr, E.~V. Sullivan, and A.~Pfefferbaum, ``The sri24 multichannel atlas of normal adult human brain structure,'' \emph{Human brain mapping}, vol.~31, no.~5, pp. 798--819, 2010.

\bibitem{kingma2014adam}
Kingma and Ba, ``Adam: A method for stochastic optimization,'' \emph{arXiv preprint arXiv:1412.6980}, 2014.

\bibitem{pmid8598820}
H.~Gudbjartsson and S.~Patz, ``{{T}he {R}ician distribution of noisy {M}{R}{I} data},'' \emph{Magn Reson Med}, vol.~34, no.~6, pp. 910--914, Dec 1995.

\end{thebibliography}
\end{document}